\documentclass[aps,pra,twocolumn,showpacs,superscriptaddress]{revtex4}
\usepackage{amsmath}
\usepackage{mathrsfs}
\usepackage{amssymb}
\usepackage{amsfonts}
\usepackage{graphicx}

\begin{document}

\title{An alternative quantum fidelity for mixed states of qudits}
\author{Xiaoguang Wang}
\email{xgwang@zimp.zju.edu.cn}

\affiliation{Zhejiang Institute of Modern Physics, Department of
Physics, Zhejiang University, HangZhou 310027, P.R. China.}
\affiliation{Department of Physics and Institute of Theoretical
Physics, The Chinese University of Hong Kong, Hong Kong.}
\author{Chang-Shui Yu}
\affiliation{School of Physics and Optoelectronic Technology, Dalian University of
Technology, Dalian 116024, P. R. China.}
\author{X. X. Yi}
\affiliation{School of Physics and Optoelectronic Technology, Dalian University of
Technology, Dalian 116024, P. R. China.}

\date{\today }

\begin{abstract}
We {give an alternative }definition of quantum fidelity for two
density operators on qudits in terms of the Hilbert-Schmidt inner
product between them and their purity. It can be regarded as the
well-defined operator fidelity for the two operators and satisfies
all Jozsa's four axioms up to a normalization factor. One desire
property is that it is not computationally demanding.
\end{abstract}

\pacs{05.45.Mt; 03.65.Nk,03.65.Yz}
\maketitle

Fidelity is an important concept in quantum information theory~\cite{Nielsen}
and quantum chaos~\cite{fqchaos1}. The well-known quantum fidelity for two
general mixed states $\rho_0$ and $\rho_1$ is given by the Uhlmann's fidelity~\cite%
{Uhlmann,Hubner,Jozsa, Schumacher}
\begin{align}
\mathcal{F}\left(\rho_0,\rho_1\right) =\text{tr}\sqrt{\rho_0^{1/2}{\rho_1}%
\rho_0^{1/2}}.
\end{align}%
This fidelity has many nice properties such as concavity and
multiplicativity under tensor product and it satisfies all Josza's
four axioms~\cite{Jozsa}. {However, it is not an easy task to make
analytical evaluation of the fidelity and even numerical
calculations due to the square roots of Hermitian matrix in the
above equation}. Quite recently, people tried
to define new fidelities to avoid this difficulty. Miszczak et al. and Mendon%
\c{c}a et al. defined the following fidelity~\cite{Miszczak,Mendonca}
\begin{align}
\mathcal{F}_1\left(\rho_0,\rho_1\right) =\text{tr}(\rho_0\rho_1)+\sqrt{1-%
\text{Tr}(\rho_0^2)}\sqrt{1-\text{Tr}(\rho_1^2)},
\end{align}%
Another fidelity which is essentially is the same as $\mathcal{F}_1$ is
defined by Chen et al. as~\cite{Chen}
\begin{align}
\mathcal{F}_2\left(\rho_0,\rho_1\right)=\frac{1-r}2+\frac{1+r}2 \mathcal{F}%
_1\left(\rho_0,\rho_1\right),
\end{align}
where $r=1/(d-1)$ with $d$ being the dimension of the Hilbert space.
This fidelity displays a nice property that it has a clear
hyperbolic geometric interpretation. Another property is that these
two fidelities reduce to Uhlmann's fidelity in the special case of
dimension $d=2$.

One fundamental requirement for a definition of fidelity is that it
must obey $F(\rho,\rho)=1$. All the above three definitions
satisfies this condition. However, it is not sufficiently emphasized
in earlier studies that when two density matrix are orthogonal, the
fidelity should be zero. This could be another fundamental
requirement for the fidelity. Consider the following density
matrices
\begin{align}
\rho_0=\frac{1}2(|0\rangle\langle 0|+|1\rangle\langle 1|),  \notag \\
\rho_1=\frac{1}2(|2\rangle\langle 2|+|3\rangle\langle 3|),
\end{align}
acting on four-dimensional Hilbert space spanned by $\{|n\rangle,
n=0,1,2,3\} $. Obviously, these two density matrix is orthogonal and
the fidelity should be zero. After some simple calculations, we find
that $\mathcal{F}=0, \mathcal{F}_1=1/2$, and $\mathcal{F}_2=2/3$.
Thus, in this strict sense it is appropriate to call $\mathcal{F}_1$
super-fidelity which acts as a useful upper bound for the Uhlmann's
fidelity~\cite{Miszczak}.

In this paper, we introduce an alternative fidelity defined, which
satisfies Josza's axioms up to a normalization factor. And the
fidelity is zero when two density matrices are orthogonal and is 1
when they are identical. We also discuss its other properties such
as convexity and multiplicativity under tensor products.

The fidelity can be regarded as the operator fidelity~\cite{wang07}
and thus we begin by introducing the definition of operator fidelity
between two operators. Let $\mathcal{H}$ be a $d$-dimensional
Hilbert space. All linear
operators on $\mathcal{H}$ on its own is a $d^2$-dimensional Hilbert space $%
\mathcal{H}_{\text{HS}}$. The inner product in this space is defined as the
Hilbert-Schmidt product, i,e., for operators $A$ and $B$, $\langle
A|B\rangle=\text{Tr}(A^\dagger B)$. Thus, any linear operators on $\mathcal{H%
}$ can be considered as a state on $\mathcal{H}_{\text{HS}}$. Thus, the
fidelity of two states can be naturally be lifted to the operator level.

To define the operator fidelity between two operators $A$ and $B$, we need
to first normalize them as $A/\sqrt{\text{Tr}(AA^\dagger)}$ and $B/\sqrt{\text{Tr}%
(BB^\dagger)}$, respectively. Then, the operator fidelity is defined
as
\begin{equation}
F(A,B)=\frac{\left\vert \text{Tr}(A^{\dagger }B)\right\vert }{\sqrt{\text{Tr}%
(AA^{\dagger })\text{Tr}(BB^{\dag })}}.
\end{equation}%
If we consider two unitary operators $U_{0}$ and $U_{1}$, the above fidelity
reduces to
\begin{equation}
F(U_{0},U_{1})=\frac{1}{d}|\text{Tr}(U_{0}^{\dagger }U_{1})|,  \label{fff}
\end{equation}%
which is studied in Ref.~\cite{wang07} and can be applied to measure the
sensitivity of quantum systems to perturbations. If two density operator $%
\rho $ and $\sigma $ are considered, the operator fidelity reduces to
\begin{equation}
F(\rho_0,\rho_1 )=\frac{\left\vert \text{Tr}(\rho_0 \rho_1 )\right\vert }{\sqrt{%
\text{Tr}(\rho_0^{2})\text{Tr}(\rho_1 ^{2})}}.  \label{newfi}
\end{equation}%
This is a function of the Hilbert-Schmidt inner product and two
purity (equivalent to linear entropy). The fidelity for two density
operators can be considered as operator fidelity. On the other hand,
it can also be regarded as fidelity between two states $\rho _{0}$
and $\rho _{1}$. This is the alternative definition of the fidelity.
{One
cannot simply define the fidelity as }$|\text{Tr}(\rho _{0}\rho _{1})|${%
\ as it becomes less than one when the two density matrices are identical,
i.e., }$\text{Tr}(\rho _{0}^{2})<1$.

It is easy to show that the fidelity $F$ has the following desirable
properties:

(1) $F$ is normalized. The maximum 1 is attained if and only if
$\rho_0 =\rho_1.$

(2) $F$ is symmetric under swapping $\rho_0 \ $and $\rho_1$, i.e.,
$F(\rho_0 ,\rho_1 )=F(\rho_0 ,\rho_1).$

(3) The fidelity is invariant under unitary transformation $U$ on
the state space, i.e., $F(\rho_0 ,\rho_1 )=F(U\rho_0 U^{\dagger
},U\rho_1 U^{\dagger })$.

(4) When one of the state is pure, say,
$\rho_1=|\psi\rangle\langle\psi|$,
the fidelity reduces to $F(\rho_0,|\psi\rangle\langle\psi|)=\langle\psi|\rho_0|%
\psi\rangle/\text{Tr}(\rho_0^2)$.

To compare with Jozsa's four axioms, only the fourth property
differs by a normalization factor $1/\text{Tr}(\rho_0^2)$. Then, we
see that this fidelity satisfies all Jozsa's axioms up to a
normalization factor~\cite{Jozsa}. Another obvious fact is that if
two density matrices are orthogonal, the fidelity is zero. It is
easy to check another nice property that the fidelity $F$ is
multiplicative under tensor products, i.e., $F(\rho _{1}\otimes \rho
_{2},\sigma _{1}\otimes \sigma _{2})=F(\rho _{1},\sigma _{1})F(\rho
_{2},\sigma _{2})$. The Uhlmann's fidelity also satisfies this
property.

Next, we check that if $F$ satisfies the property of concavity or convexity.
By numerical calculations, we find that the following inequality
\begin{equation*}
p_{1}F(\rho _{1},\sigma )+p_{2}F(\rho _{2},\sigma )\leq F(p_{1}\rho
_{1}+p_{2}\rho _{2},\sigma ),p_{1},p_{2}\geq 0
\end{equation*}%
is satisfied for most matrices $\rho _{1},\rho _{2}$ and $\sigma $ for $%
p_{1}+p_{2}=1$. {However, the violation can also be found. The most
simplest demonstration can be given if }$\rho _{1}=I/2 $, $\rho
_{2}=|0\rangle\langle 0|$ and $\sigma =|1\rangle\langle 1| $ for $%
p_{1},p_{2}\in (0,1)$. Here, states $|0\rangle$ and $|1\rangle$ are
orthogonal and $I$ is the $2\times 2$ identity matrix.

To show the violation of concavity, let
\begin{equation*}
\tilde{\rho}=p\rho _{1}+(1-p)\rho _{2}=(1-p/2)|0\rangle \langle
0|+p/2|1\rangle \langle 1|,
\end{equation*}%
then
\begin{equation*}
F(\tilde{\rho},\sigma )=\frac{\langle 1|\tilde{\rho}|1\rangle }{\sqrt{\text{%
Tr}(\tilde{\rho}^{2})}}=\frac{p}{\sqrt{2}[1+(1-p)^{2}]^{1/2}},
\end{equation*}%
and
\begin{equation*}
pF(\rho _{1},\sigma )+(1-p)F(\rho _{2},\sigma )=pF(\rho _{1},\sigma )=\frac{p%
}{\sqrt{2}}.
\end{equation*}%
It is obvious that
\begin{equation*}
F(\tilde{\rho},\sigma )<pF(\rho _{1},\sigma )+(1-p)F(\rho
_{2},\sigma )
\end{equation*}%
for $p\in (0,1)$.

That is to say, $F$ satisfies neither concavity nor convexity. Since
any measure is monotonically increasing (decreasing) if it is (i)
unitarily invariant, (ii) jointly concave (convex) and (iii)
invariant under the addition of an ancillary system~\cite{Lieb}, $F$
is not monotonically increasing or decreasing under quantum
operations.

As an application, we consider thermal equilibrium density matrix $%
\rho_{k}=\exp(-\beta H_k)/Z(\beta) (k=0,1)$ acting on $d$-dimensional
Hilbert space [$Z_k(\beta)=\text{Tr}[\exp(-\beta H_k)]$ is the partition
function for $k$-th system, $T=\beta^{-1}$ is the temperature, and the
Boltzmann constant is assumed to be one]. From Eq.~(\ref{newfi}), the
fidelity for the two thermal states is given by
\begin{align}  \label{newthermal}
&F(\rho_0,\rho_1)  \notag \\
&=\frac{\text{Tr}(e^{-\beta H_0}e^{-\beta H_1})}{\sqrt{\text{Tr}(e^{-2\beta
H_0})\text{Tr}(e^{-2\beta H_0})}}  \notag \\
&=\frac{\text{Tr}[(e^{-\beta H_0})^\dagger e^{-\beta H_1}]}{\sqrt{\text{Tr}%
[(e^{-\beta H_0})^\dagger e^{-\beta H_0}]\text{Tr}[(e^{-\beta H_1})^\dagger
e^{-\beta H_1}]}}.
\end{align}
It is well-known that imaginary time (or imaginary temperature) is essential
in connecting quantum mechanics and statistical mechanics. If we make the
Wick rotation, i.e., let $\beta=it$, the above equation reduces to
\begin{equation}  \label{newthermal1}
F(U_0,U_1)=1/d {\left|\text{Tr}(e^{itH_0}e^{-it H_1})\right|},
\end{equation}
which is just the operator fidelity for two unitary operators $U_k$
generated by Hamiltonian $H_k$. Thus, we see that the fidelity for
two thermal states is connected to the operator fidelity for two
unitary evolution operators by the Wick rotation by $\pi/2$. The
fidelity introduced here is expected to be applicable to studies of
phase transitions and quantum chaos.

In conclusion, we have introduced an alternative fidelity which
satisfiies Jozsa's four axioms up to an normalization factor. It has
a desire property that is multiplicative under tensor products and
undesire one that it is neither convex nor concave. The relations
between this fidelity and the operator fidelity was clarified.
Another merit is that it is not computationally demanding. From an
measurement point of view, this fidelity is relatively easy to
measure as it contains only the Hilbert-Schmidt inner product and
two purity.

\section{Acknowledgements}

We are indebted to S. J. Gu and Z. W. Zhou for fruitful and valuable
discussions. X. Wang acknowledges the support from C. N. Yang
foundation of CUHK, the Program for New Century Excellent Talents in
University (NCET), the NSFC with grant nos. 90503003, the State Key
Program for Basic Research of China with grant nos. 2006CB921206,
and the Specialized Research Fund for the Doctoral Program of Higher
Education with grant No.20050335087.

\end{document}